\documentclass{aa}
\newcommand{\degs}{$^{\circ}$}
\newcommand{\pam}{.\hskip-2pt$^\prime$}

\usepackage[dvips]{graphics}
\usepackage{psfig}

\def\B40{BD$+40^\circ 4124$}
\def\bd40{BD$+40^\circ 4124$}
\def\BD61{BD$+61^\circ 154$}
\def\bd61{BD$+61^\circ 154$}

\def\hdo {H$_2$O}

\newcommand{\kms}{km~s{$^{-1}$}}

\newcommand{\hii}{H{\sc ii}}

\begin{document}
\title{Water maser variability over 20 years in a large sample of 
star-forming regions: the complete database\thanks{Based on observations 
with the Medicina radiotelescope operated by INAF -
Istituto di Radioastronomia}}

\author{M. Felli\inst{1} \and J. Brand\inst{2} \and
R. Cesaroni\inst{1}  \and C. Codella\inst{3} \and  G. Comoretto\inst{1} \and 
S. Di Franco\inst{4} \and F. Massi\inst{1} \and L. Moscadelli\inst{1} \and R. Nesti\inst{1}
\and L. Olmi\inst{3} \and
F. Palagi\inst{3} \and D. Panella\inst{1} \and R. Valdettaro\inst{1} 
}
\institute{
INAF-Osservatorio Astrofisico di Arcetri, Largo E. Fermi, 5, I-50125 Firenze, Italy
\and
INAF-Istituto di Radioastronomia, Via Gobetti 101, I-40129 Bologna, Italy
\and
INAF-Istituto di Radioastronomia, Sez. di Firenze,
Largo E. Fermi 5, I-50125 Firenze, Italy
\and
Dipartimento di Astronomia, Universit\`a degli Studi di Firenze,
Largo E. Fermi 2, I-50125 Firenze, Italy}

\offprints{R. Cesaroni, \email{cesa@arcetri.astro.it}}
\date{Received; Accepted }

\titlerunning{Variability of \hdo\  masers in SFR}
\authorrunning{M. Felli et al.}

\abstract{Water vapor emission at 22 GHz from masers associated with 
star-forming regions is highly
variable.}{We present  a database of up to 20  years of monitoring 
of a sample of 43 masers within star-forming regions.  The sample covers a large range of
luminosities of the associated IRAS source and is representative of the
entire population of \hdo\ masers of this type.  The database forms a good starting point
for any further study of \hdo\ maser variability.}{The 
observations were obtained with the
Medicina 32--m radiotelescope, at  a rate of 4--5 observations per year.} 
{To provide a
database that can be easily  accessed through the web,  we give
for each source: plots of the calibrated spectra, the velocity--time--flux density plot, 
the light curve of the integrated flux, the lower and upper envelopes of the maser emission, the
mean spectrum, and 
the rate of the maser occurrence as a function of velocity.
Figures for just  one source  are given in the text for representative purposes. Figures 
for all the sources are given in electronic form in the on-line appendix.
A discussion of the main
properties of the \hdo\ variability in our sample
will be presented in a forthcoming paper.}
{}

\maketitle

\section{Introduction}

Since the discovery of 22 GHz \hdo\ maser emission 
associated with young stellar
objects (YSOs) within star-forming regions (SFRs), variability of
maser emission is well-known.  Changes 
as large as several orders of magnitude in the maser emission 
have been observed
(e.g., Little et al.~\cite{lit77}; 
Liljestr\"om et al.~\cite{lil89}; Claussen et al.~\cite{cla96};
Comoretto et al.~\cite{com90}; Wouterloot et al.~\cite{wou95}). At the same time,  velocity drifts of
individual components of up to a few \kms\ per year  have also 
been reported (e.g.,  Brand et al.~\cite{BCCFPPV2003}).
The variability can be slow or burst-like and covers all ranges of timescales,  
from hours-days to months-years.
In the present study we are mainly concerned with the latter and, for the first time,
deal with a large sample of sources (43) and with a time-span of up to 20 years,
with 4--5 spectra per year.

Due to the large amount of  telescope time required to follow the evolution
of \hdo\ maser  emission, inevitably the variability is more easily  monitored through single dish
observations. 
In fact, besides our campaign only the Pushchino 22--m single-dish
   maser patrol covers a comparably long period (e.g., Rudnitskij
et al.~\cite{russi}).
%
%
Furuya et al.\ (\cite{furuya}) carried out a multiepoch 22 GHz H$_{2}$O
maser survey towards 173 low-mass YSOs (Class 0 to Class III sources) 
using the Nobeyama 45 m telescope. This was the first complete water
maser survey towards Class 0 sources in the northern sky.
However, their observations extend non-uniformly over
a period of only three years.
%
%

It is well-known that
the variability depends strongly  on the luminosity of the  SFR (usually
derived from the associated IRAS source).  For instance,
there is a minimum luminosity ($\sim$25 L$_\odot$)  below which
\hdo\ maser emission may be present only for about one third of the entire duration
of the maser activity (Persi et al.~\cite{per94}; Claussen et al.~\cite{cla96}).
The results obtained so far suggest that the lower
the SFR luminosity, the higher the observed degree of variability
of the \hdo\ maser emission.
Higher luminosity sources may show more steady components.
Consequently, 
a large  sample of sources is  needed
to better understand the dependence of the variability 
on other parameters of the SFRs, in particular on  their luminosities which cover a range
of several orders of magnitude, and on their (molecular) environment. 

The use of a single dish instrument is of course
a limitation because interferometric observations clearly show that 
\hdo\  masers within a SFR 
often consist of  many spatially separated, unresolved 
components, generally clustered in groups (usually with different
velocities), which in most cases cannot be separated in 
single dish observations (e.g., Forster \& Caswell~\cite{for89}, 
~\cite{for99}; Tofani et al.~\cite{tof95}). 
However, Felli et al.~(\cite{FMRC2006}) reported a case
in which single dish observations were able to separate the evolution
of spatially distinct components.
Nevertheless, frequent single dish observations have the advantage 
of being able to follow potential  velocity drifts of individual
{\it velocity} components, which can be easily  identified in the
spectra, and thus give a better view of the dynamics (e.g., accelerations)
occurring in the circumstellar environment of the YSO. 

In two earlier papers (Valdettaro et al.~\cite{VPBCCFP2002};
Brand et al.~\cite{BCCFPPV2003}), we set the basis for a systematic study of the \hdo\ variability
in SFRs with the Medicina 32--m radiotelescope  using a sample of 14 sources which  covered  
a wide range of luminosities and
had been observed  for more than 10 years. 
Here we report on a larger sample of 43 SFRs (including the former 14 
sources)  and increase  the time coverage up to 20 years. 
The database is presented in the form of plots of the calibrated spectra and
additional plots of derived quantities, in a form that can be easily accessed through  the web. 
In a forthcoming paper, we will analyze  the properties
of \hdo\ variability of our  sample.

\section{The sample}
\label{sres}

A large sample of sources selected from the Arcetri Atlas (Cesaroni et al.~\cite{cesa88};
Comoretto et al.~\cite{com90}; Palagi et
al.~\cite{PCC93}; Brand et al.~\cite{brand}; Valdettaro et al.~\cite{val01}) has
been monitored regularly since the beginning of the operating period of the
Arcetri digital autocorrelator at the Medicina 32--m radiotelescope (1987), at a
rate of 4--5 observations per year.
Our goal was 
to study the dependence of the variability of \hdo\ masers in  SFRs on the
luminosity of the associated YSO, assumed to remain constant during our patrol.
It should be pointed out that the FIR luminosity is derived from IRAS observations 
and, consequently, is representative of the {\it entire} luminosity of the SFR and not necessarily
of the YSO that powers the \hdo\ maser. In fact a SFR may have 
a very complex structure, with  several YSOs in various  evolutionary stages and
with different luminosities  present within the same  SFR,
as for instance in W75--N (Hunter et al.~\cite{hun94}).  
Variability might  also depend on other characteristics of the SFR such as the 
presence of molecular outflows (Felli et al.~\cite{FPT92}), its
IRAS colours and the presence or absence of
an UC \hii\ region (Palla et al.~\cite{pal91},~\cite{PCBCCF93};
Codella et al.~\cite{CFNPP94},~\cite{CFN96};
Codella \& Felli~\cite{CF95}; Codella \& Palla~\cite{CP95}).

\begin{figure}[tp]
\resizebox{9cm}{!}{ \rotatebox{0}{\includegraphics{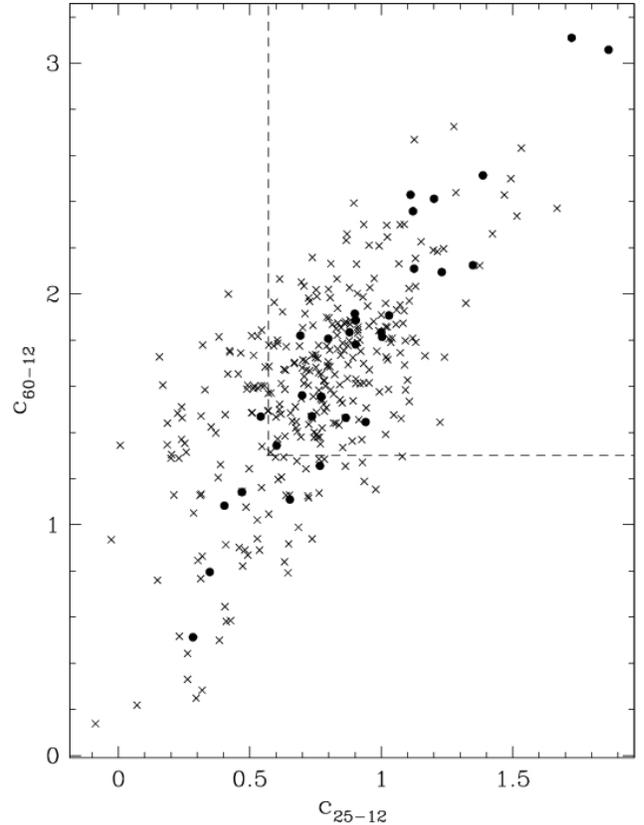} }}
\caption[]{Colour-colour (C$_{a-b}$ = log(F$_a$/F$_b$), F$_a$ = IRAS flux
in band $a$ $\mu$m) diagram of the IRAS counterparts of the sources
listed in Table~1 ($\circ$), compared to that of a larger sample of \hdo\ masers of SFR type
(a merger of Comoretto et al.~\cite{com90},
Brand et al.~\cite{brand} and Valdettaro et al.~\cite{val01}) 
($\times$). The dashed box defines the location  
of Ultracompact \hii\
regions (Wood \& Churchwell~\cite{WC89})}
\label{var}
\end{figure}

The present sample of 43 \hdo\ masers was chosen out of
the several hundred  known \hdo\ masers of SFR 
type (e.g., Comoretto et al.~\cite{com90}) using the following criteria:
\begin{itemize}
\item [(1)]
the maser sources were to be of the SFR-type, i.e. associated with a
SFR, to be distinguished from those associated with late-type stars. 
The SFR classification  of Palagi et
al.~(\cite{PCC93}) and Valdettaro et al.~(\cite{val01}) is based on IRAS colours.
The location  of the sources of our sample in an IRAS colour-colour plot
is shown in Fig.~\ref{var} and compared with that of a much larger sample
of SFRs with \hdo\ maser emission (a merger of Comoretto et al.~\cite{com90},
Brand et al.~\cite{brand} and Valdettaro et al.~\cite{val01}).
The dashed box  defines the region
occupied by sources with colours characteristic of ultracompact \hii\ regions
(Wood \& Churchwell~\cite{WC89}).
To double-check the true nature of the sources exciting the \hdo\ masers,
other morphological indicators of the SFRs were also searched for in the literature,
e.g.
association with dense molecular cores (e.g Cesaroni et al.~\cite{CFW98}),
presence of UC \hii\ regions,
presence of molecular outflows and other types of masers. All investigations  further
support the SFR identification\footnote{
We note that in SIMBAD, AFGL2789 is identified with ``V645 Cyg, pulsating variable star''.
However, as stated by Clarke et al.~(\cite{clarke}),
V645 Cyg is a ``relatively unembedded young massive star, with a high-velocity
wind and an associated optical and molecular outflow'', and it ``may represent
a relatively rare class of transition objects between a genuine massive YSO and a normal young
Oe-type star in a weak \hii\ region''. Its presence in our sample is therefore justified.}.
\item [(2)]
the luminosities of the associated IRAS source was to cover as large a range  as possible; 
\item [(3)] the sample was to be  sufficiently large
to be representative of the entire population of \hdo\ masers within SFRs;
\item [(4)]
the sources were to be suitably distributed in right ascension, to avoid 
scheduling problems due to clustering
of sources in the inner region of the Galactic plane in the large sample; 
\item [(5)]
the peak flux density was to be greater than a  few Jy 
so that the source could be easily observed with a 5 min 
on-source integration time with the Medicina 32--m radio telescope;
\item [(6)] 
the number of selected sources was not to be too large to allow observations  in a
$<$ 2--3 day session;
\item [(7)] finally, in order to prevent confusion in the single beam from nearby,
but unrelated, maser sources,
we selected sources from the Arcetri Atlas for which no
other masers had been reported within  a circle of several
arcminutes with single dish observations\footnote{
A possible exception might be OMC 2 and KL IRC2 which are
separated by 13.\arcmin 3.  During the
mega burst of KL IRC2 in January 1999, in which the flux
density increased by at least a factor 10$^3$ and reached $\sim$ 2.6 10$^7$ Jy, we observed
a similar burst in OMC 2 at the same velocity  (around 8 \kms) and with
a flux density of $\sim$ 3.6 10$^4$ Jy.
Beam shape mapping  indicates that at this distance
from the beam axis the  attenuation should be about 10$^{-3}$.
Consequently,  a side-lobe effect of the observed intensity is
expected. However, apart for the period of the mega-burst, all the remaining
data for OMC 2 should be unaffected by the presence of the
close-by source KL IRC2, at least outside the velocity range close to the KL IRC2
peak emission, approximately  for velocities less than 6 \kms or greater than 10 \kms. In fact,
1) when the flux of KL IRC2 is at a level of  
2--3 10$^3$ Jy there are OMC 2 spectra with no signal above
the noise level of 1--2 Jy, 2) there are features observed in OMC 2  
outside the above-mentioned velocity range which do not match with any emission
in KL IRC 2.
In conclusion, the data of OMC 2 should be used with the caveat of possible 
KL IRC2 side lobe effects in the range of velocities with strong KL IRC2 emission.}.

\end{itemize}

The main source parameters are given in
Table~\ref{tsample}, which lists: 
\begin{itemize}
\item [(1)] a running  number;
\item [(2)] the source name; 
\item [(3)] the associated IRAS source, when available;
\item [(4)] J2000.0 right ascension; 
\item [(5)] J2000.0 declination;
\item [(6)] $V_{\rm cl}$, the  velocity of the
associated molecular cloud relative to the LSR, derived from
the literature using tracers sensitive to high molecular density, such as NH$_3$ and
CS (in \kms)
\item [(7)] $d$, the distance 
taken from the literature (in kpc); 
\item [(8)] $L_{\rm IR}$, the  integrated FIR luminosity (in L$_{\odot}$), 
usually derived from IRAS 
data or, otherwise, taken from the literature; 
\item [(9)] the date of the first observation; 
\item [(10)] references for the distance.
\end{itemize}

The IRAS luminosity was obtained  
by adopting the distance from Col.~7, assuming an emissivity proportional
to the frequency, correcting the observed IRAS flux densities for the derived colour
temperatures between two adjacent bands, and extrapolating the fluxes to 6
and 400~$\mu\mathrm{m}$ (Wouterloot et al.~\cite{wou95}).

%
\begin{table*}
\caption{The \hdo\ maser sample}
\begin{tabular}{c c c c c c c c c c}     

\hline
\# & Name & IRAS & $\alpha$(J2000) & $\delta$(J2000) & $ V_{\rm cl}$ & $d$ & $L_{\rm IR}$ & First obs. & Ref. \\
   &      &      &                 &                 & (km s$^{-1}$) & (kpc) & ($L_{\sun}$) &                 & \\
\hline
  1 & \object{NGC 281}         & 00494+5617 & 00:52:24.7 & +56:33:50 & --30.8  & 2.2  &   7.9~10$^{3}$ & 01-APR-1987 & 1 \\
  2 & \object{W3 OH}           & 02232+6138 & 02:27:04.7 & +61:52:26 & --46.8  & 2.04  &  6.6~10$^{4}$ & 27-MAR-1987 & 2 \\
  3 & \object{RNO 15--FIR}     & 03245+3002 & 03:27:38.1 & +30:12:59 &   4.7  & 0.35  &  2.0~10$^{1}$ & 30-MAR-1987 & 3 \\
  4 & \object{AFGL 5142}       & 05274+3345 & 05:30:48.0 & +33:47:54 &  --4.1  & 1.8  &   5.3~10$^{3}$ & 14-SEP-1995 & 4 \\
  5 & \object{Ori A--west}     & 05302-0537 & 05:32:41.6 & --05:35:47 &   8.9  & 0.5  &   4.0~10$^{1}$ & 13-FEB-1990 & 5,6 \\
  6 & \object{KL IRC2}         &            & 05:35:14.5 & --05:22:30 &   9.0  & 0.45  &  8.0~10$^{4}$ & 30-MAR-1987 & 6 \\
  7 & \object{OMC 2}           & 05329-0512 & 05:35:27.6 & --05:09:35 &  11.1  & 0.45  &  2.0~10$^{2}$ & 26-MAR-1987 & 6 \\
  8 & \object{Sh 2--231}       & 05358+3543 & 05:39:12.9 & +35:45:51 & --17.6  & 2.00  &  6.4~10$^{3}$ & 23-MAR-1987 & 7 \\
  9 & \object{HHL 26}          & 05373+2349 & 05:40:24.1 & +23:50:55 &   2.0  & 2.4  &   1.8~10$^{3}$ & 09-JUL-1988 & 8 \\
 10 & \object{Sh 2--235}       & 05375+3540 & 05:40:53.3 & +35:41:49 & --57.0  & 1.8  &   1.1~10$^{4}$ & 31-MAR-1987 & 9,34 \\
 11 & \object{NGC 2071}        & 05445+0020 & 05:47:05.4 & +00:21:43 &   9.5  & 0.39  &  4.1~10$^{2}$ & 20-MAR-1987 & 10 \\
 12 & \object{HH 397A}         & 05553+1631 & 05:58:13.9 & +16:32:00 &   5.7  & 2.0  &   3.0~10$^{3}$ & 20-MAR-1989 & 11 \\
 13 & \object{Mon R2 IRS3}     & 06053-0622 & 06:07:48.0 & --06:22:57 &  10.5  & 0.8  &   3.2~10$^{4}$ & 03-MAY-1988 & 12 \\
 14 & \object{Sh 2--252}       & 06055+2039 & 06:08:35.5 & +20:39:13 &   8.9  & 2.0  &   4.0~10$^{3}$ & 01-APR-1987 & 13 \\
 15 & \object{AFGL 5180}       & 06058+2138 & 06:08:53.3 & +21:38:12 &   3.3  & 1.5  &   5.0~10$^{3}$ & 19-DEC-1998 & 4 \\
 16 & \object{GGD 12--15}      & 06084-0611 & 06:10:52.2 & --06:11:32 &  11.4  & 1.0  &   7.9~10$^{3}$ & 17-JUN-1987 & 14 \\
 17 & \object{Sh 2--255/7}     & 06099+1800 & 06:12:53.6 & +17:59:27 &   7.0  & 2.5  &   4.7~10$^{4}$ & 22-MAR-1987 & 15 \\
 18 & \object{Sh 2--269}       & 06117+1350 & 06:14:36.5 & +13:49:35 &  18.2  & 5.3  &   1.2~10$^{5}$ & 22-MAR-1987 & 16 \\
 19 & \object{NGC 2264}        & 06384+0932 & 06:41:10.1 & +09:29:22 &   9.0  & 0.8  &   2.1~10$^{3}$ & 24-OCT-1992 & 17 \\
 20 & \object{G31.41+0.31}     & 18449-0115 & 18:47:34.7 & --01:12:46 &  98.0  & 7.9  &   2.6~10$^{5}$ & 22-MAR-1987 & 18 \\
 21 & \object{W43 Main3}       &            & 18:47:47.0 & --01:54:35 &  98.7  & 6.54  &  1.5~10$^{6}$ & 31-MAR-1987 & 19 \\
 22 & \object{G32.74--0.08}    & 18487-0015 & 18:51:21.9 & --00:12:09 &  38.2  & 2.6  &   5.3~10$^{3}$ & 12-JUN-1987 & 20 \\
 23 & \object{G34.26+0.15}     & 18507+0110 & 18:53:18.8 & +01:14:56 &  57.8  & 3.9  &   7.5~10$^{5}$ & 22-MAR-1987 & 20,18 \\
 24 & \object{G35.20--0.74}    & 18556+0136 & 18:58:12.6 & +01:40:37 &  34.0  & 1.8  &   1.4~10$^{4}$ & 01-APR-1987 & 20 \\
 25 & \object{OH43.8--0.1}     & 19095+0930 & 19:11:54.2 & +09:35:55 &  41.0  & 2.8   &  2.7~10$^{4}$ & 22-MAR-1987 & 21 \\
 26 & \object{G45.07+0.13}     & 19110+1045 & 19:13:22.0 & +10:50:52 &  58.5  & 6.0  &   4.4~10$^{5}$ & 01-APR-1987 & 18 \\
 27 & \object{G59.78+0.06}     & 19410+2336 & 19:43:11.5 & +23:43:54 &  22.3  & 2.2  &   1.5~10$^{4}$ & 01-APR-1987 & 22 \\
 28 & \object{ON 1}            & 20081+3122 & 20:10:09.1 & +31:31:37 &  13.0  & 3.0  &   2.7~10$^{4}$ & 23-MAR-1987 & 7 \\
 29 & \object{IRAS 20126+4104} & 20126+4104 & 20:14:26.0 & +41:13:33 &  --3.6  & 1.7  &   1.0~10$^{4}$ & 21-MAR-1989 & 23 \\
 30 & \object{AFGL 2591}       & 20275+4001 & 20:29:24.9 & +40:11:20 &  --5.7  & 1.0  &   2.0~10$^{4}$ & 31-MAR-1987 & 24 \\
 31 & \object{W75--N}          &            & 20:38:36.4 & +42:37:35 &  10.0  & 2.0  &   1.4~10$^{5}$ & 22-MAR-1987 & 25 \\
 32 & \object{Sh 2--128(H$_2$O)} & 21306+5540 & 21:32:11.4 & +55:53:55 & --71.0  & 6.5  &   8.9~10$^{4}$ & 26-MAR-1987 & 20 \\
 33 & \object{AFGL 2789}       & 21381+5000 & 21:39:58.2 & +50:14:22 & --43.9  & 5.7  &   4.5~10$^{4}$ & 29-JAN-1989 & 26 \\
 34 & \object{IC1396n}         & 21391+5802 & 21:40:41.9 & +58:16:12 &   0.0  & 0.62  &  3.2~10$^{2}$ & 21-MAR-1989 & 27 \\
 35 & \object{NGC 7129 FIRS2}  &            & 21:43:00.2 & +66:03:26 & --10.1  & 1.0  &   4.3~10$^{2}$ & 26-OCT-1991 & 20 \\
 36 & \object{Sh 2--140 IRS1}  & 22176+6303 & 22:19:18.3 & +63:18:47 &  --7.1  & 0.9  &   2.6~10$^{4}$ & 30-MAR-1987 & 7 \\
 37 & \object{L1204--G}        & 22198+6336 & 22:21:26.7 & +63:51:38 & --10.8  & 0.9  &   5.9~10$^{2}$ & 04-DEC-1989 & 28,7 \\
 38 & \object{IRAS 22506+5944} & 22506+5944 & 22:52:36.9 & +60:00:48 & --51.5  & 5.7  &   2.2~10$^{4}$ & 12-JUN-1987 & 29 \\
 39 & \object{Cepheus A}       & 22543+6145 & 22:56:18.1 & +62:01:46 & --10.7  & 0.73  &  2.5~10$^{4}$ & 11-FEB-1987 & 30 \\
 40 & \object{WB89--234H$_2$O} & 23004+5642 & 23:02:31.8 & +56:57:44 & --53.5  & 5.6  &   9.6~10$^{3} $ & 17-SEP-1995 & 31 \\
 41 & \object{Sh 2--158}       & 23116+6111 & 23:13:44.7 & +61:28:10 & --56.9  & 2.5  &   2.5~10$^{5}$ & 12-JUN-1987 & 32 \\
 42 & \object{IRAS 23139+5939} & 23139+5939 & 23:16:10.3 & +59:55:29 & --44.0  & 3.5  &   1.0~10$^{4}$ & 12-JUN-1987 & 33 \\
 43 & \object{IRAS 23151+5912} & 23151+5912 & 23:17:20.8 & +59:28:47 & --54.7  & 3.5  &   3.9~10$^{4}$ & 12-JUN-1987 & 33 \\
 \hline\\
\end{tabular}

  \hspace{1mm} Distances from:  (1) Georgelin (\cite{geo75}); (2) Hachisuka et al.\ (\cite{hachi06});
 (3) Herbig \& Jones (\cite{hj83}); (4) Snell et al.\ (\cite{sn88}); (5)
 adopted distance by Fukui et al. (\cite{fukui}), Meehan et al.\
(\cite{meeha}) quote $d = 0.45$ kpc; (6) adopted distance, for a recent review
of distance determinations to the Orion region see Jeffries (\cite{jeff}); (7)
Palagi et al.\ (\cite{PCC93}); (8) Molinari et al.\ (\cite{mol02}); (9)
Evans \& Blair (\cite{nayo}); (10) Anthony-Twarog (\cite{antw}); (11)
Shepherd \& Churchwell (\cite{devi}); (12) Racine (\cite{raci}); (13) Lada \& Wooden
(\cite{lawo}); (14) Racine \& van den Bergh (\cite{ravdb}); (15) Hunter \&
Massey (\cite{massey}); (16) M. Honma, private communication; (17) Walker (\cite{walker}); (18)
Churchwell et al.\ (\cite{cwc}); (19) kinematic distance, using the rotation
curve of Brand \& Blitz (\cite{bb93}); (20) Valdettaro et al.\
(\cite{VPBCCFP2002}); (21) Honma et al.\ (\cite{hon05}); (22) Y. Xu, private communication; (23) Wilking et al.\
(\cite{wilky}); (24) van der Tak et al.\ (\cite{vdt});
(25) Hunter et al.\ (\cite{hun94}); (26) Clarke et al.
(\cite{clarke}); (27) de Zeeuw et al. (\cite{deze}); (28) Crampton \& Fisher
(\cite{cramfis}); (29) Molinari et al.\ (\cite{mol96}); (30) Blaauw et al.\
(\cite{bla59}); (31) Brand \& Wouterloot (\cite{bw98}); (32) L. Moscadelli,
private communication; (33) Wouterloot \& Walmsley (\cite{wouw}); (34) Armandroff
\& Herbst (\cite{arma}).
\label{tsample}
\end{table*}

\begin{figure}[tp]
\resizebox{9cm}{!}{ \rotatebox{0}{\includegraphics{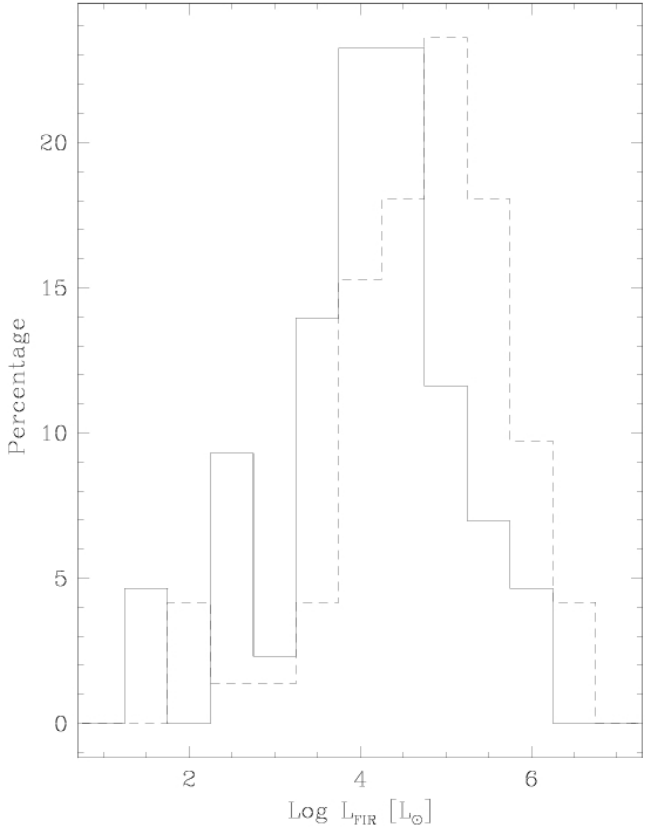} }}
\caption[]{The distribution of  IR luminosity of the SFRs in our sample (solid line) and that
of a larger sample of SFRs (Palagi et al.~\cite{PCC93}) (dashed line). Both histograms are
normalized to the total number of sources.
}
\label{lum}
\end{figure}

The sample covers a range of IR luminosities from 20~L$_{\odot}$ to
$1.5\times 10^6$~L$_{\odot}$.  These values bracket the entire luminosity
interval of the 
%
%
regions where
\hdo\ maser of the SFR-type are found
%
%
(Palagi et al.~\cite{PCC93}). 
In the histogram of Fig.~\ref{lum}  the 
distribution of the IR  luminosities of the sources in our sample is  compared with that 
of a much
larger  sample of SFR with \hdo\ maser emission (Palagi et al.~\cite{PCC93}). 
Both histograms are  normalized to their respective total number of sources.
The small decrease of sources in our sample  at large luminosities is probably due to 
the clustering of bright masers in the inner region of the Galactic plane
which is less sampled according to criterion (4). The slightly larger percentage  
of sources in our sample at low luminosities is due to \hdo\ masers
added after  the publication of the Palagi et al.~(\cite{PCC93}) sample.

\begin{figure}[tp]
\resizebox{9cm}{!}{ \rotatebox{0}{\includegraphics{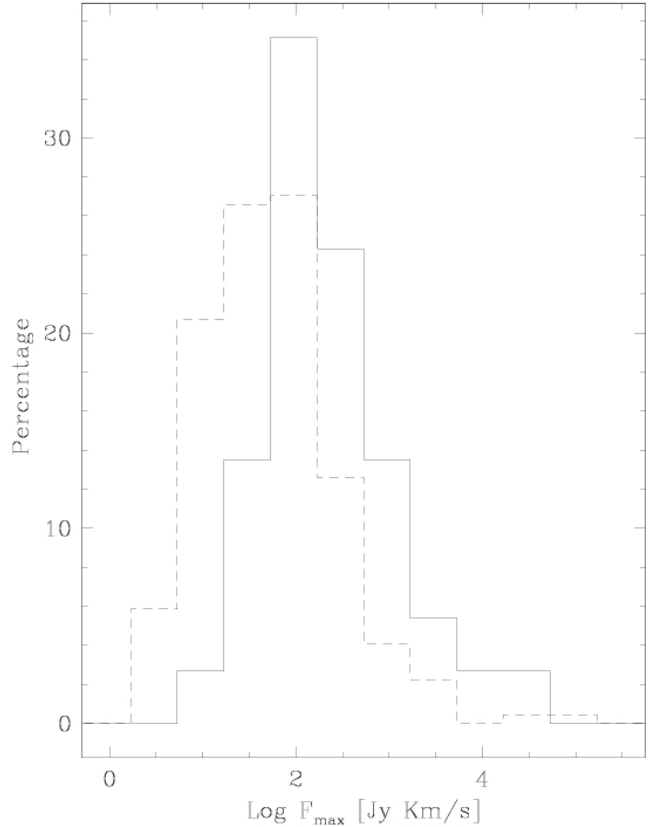} }}
\caption[]{The distribution of  the integrated fluxes of the \hdo\ 
masers in our sample as measured in the Feb.~2007 observing session
(solid line), and that 
of a larger sample of SFRs (a merger of Comoretto et al.~\cite{com90},
Brand et al.~\cite{brand} and Valdettaro et al.~\cite{val01}) (dashed line).
}
\label{flux}
\end{figure}

Similarly, in Fig.~\ref{flux}
the distribution of \hdo\ maser integrated  fluxes 
of our sources  in the last observing run (February 2007) is  compared with that of 
a much larger sample of SFR (a merger of Comoretto et al.~\cite{com90}, 
Brand et al.~\cite{brand} and Valdettaro et al.~\cite{val01}). 
Both histograms are normalized to their respective total number of sources.

Figure~\ref{flux} shows that in Feb.~2007 there were percentage-wise,
more sources with high integrated flux densities in our present sample than
there are in the comparison catalogue. Likewise there is
a smaller percentage of sources with low integrated flux densities.
For a large range in $L_{\rm IR}$ the general trend is for the average integrated
flux density to increase with increasing $L_{\rm IR}$ (e.g.,
Brand et al.~(\cite{BCCFPPV2003}); this trend will be further investigated in a
forthcoming paper) and the difference between the distributions
shown in Fig.~\ref{flux} reflects the fact that we have biased our present sample
towards sources of high $L_{\rm IR}$, in order to have a good chance of yielding a
maser detection in a reasonable amount of integration time.
On the other hand, sources of lower $L_{\rm IR}$ usually have a lower flux density,
and do not have a 100\% detection rate
(e.g., Brand et al.~\cite{brand2007}). This, in combination with our
present sample's bias towards sources with higher $L_{\rm IR}$ can be seen as the
main cause of the difference between the histograms in Fig.~\ref{flux} for the lower
integrated flux density bins. Furthermore, on the low flux density side, the difference
between the histograms is enhanced by the way sources from the
comparison catalogue were counted (if there was more than one
observation, we always selected the one in which the source was detected).
Note that although in Fig.~\ref{flux} we show the distribution of integrated flux
densities for the observing session of Feb.~2007, any other session would
have given the same statistical result.
In summary and within the limitations inherent in the comparison, we confirm that 
our sample is fairly  representative of the entire population of SFRs.

Ideally, from the observed maser variability, we would like to infer the
properties of the exciting YSO and of its surroundings. However, while for \hdo\ masers
from late--type stars the association is unambiguous (i.e.  there is only
one star pumping the maser), for SFR-type masers, several distinct YSOs may 
simultaneously be present within our beam.
This ambiguity can only be solved
with high--resolution interferometric observations, but  even if
in many cases these are available, they cover at most a few epochs and there is 
no  guarantee that the spatial/velocity situation of the maser spots remains the same
over long time intervals, given the large variability observed. 

\section{Observations}
\label{obs}

The present study is based on observations carried out with the Medicina 32--m
radiotelescope\footnote{The Medicina 32--m  VLBI radiotelescope is operated by
INAF--Istituto di Radioastronomia.}
(HPBW $\sim$1\pam 9 at 22~GHz).
This is  primarily set up 
to be used for VLBI measurements, and therefore the front-end is optimized for
such work. 

The monitoring program is ongoing but
the time interval considered in this work spans  20 years, from March
1987 to February 2007, with shorter time coverage for  a few sources which
were added to the list at a later date. 
During this period, various parts of  the whole system (antenna,
receiver, autocorrelator) were improved, increasing the
sensitivity and spectral resolution of the observations.  A detailed
description of the system and its improvements can be found in Comoretto 
et al.~(\cite{com90}) and in Brand et al.~(\cite{brand}), respectively. 
In 1989 realigning of  the antenna surface resulted in an improvement of the
efficiency at 22~GHz from 17 to 38\%. The original backend was a 512-channel
digital autocorrelator; in 1991 the number of channels was increased
to 1024. In the same year a HEMT amplifier replaced the GaAs FET
front-end, reducing the system temperature. 
In 1991 an active sub-reflector control
increased the gain at low and high elevations. 

The available bandwidth
varied between 3.125~MHz and 25~MHz (at 22~GHz this corresponds to  a
velocity resolution between 0.041~\kms\ and 0.658~\kms). In February
1997, the VLA-1 chip correlator  was replaced by a new one based on the
NFRA correlator chips and boards (Bos~\cite{bos}).  
%
%
The new autocorrelator has 2048 delay channels and 160 MHz maximum
 bandwidth in its standard configuration.
%
%
In order to maintain compatibility with
previous observations, we have continued to use only 1024 channels and a 10
MHz bandwidth, giving a velocity resolution of 0.132~\kms. 
Finally, owing to extensive maintenance works on  the
radiotelescope, there is a gap in the observations from April 1996 to
February 1997.

The telescope pointing model is typically updated a few times per year, and
is quickly checked every few weeks
by observing strong maser sources (e.g., W3~OH, Orion-KL, W49~N, Sgr~B2
and W51). The pointing accuracy was always better than 25\arcsec; from about
mid-2004 the rms residuals from the pointing model were about 
8\arcsec--10\arcsec. 

Observations were taken in position-switching mode, with both ON and OFF
scans of 5~min duration. The OFF position was taken 1.25\degr\ E of the
source position to rescan the same path as the ON scan. 
Typically, one ON/OFF pair was taken, though 
for weak sources this was repeated several times. The
typical 1 $\sigma$ noise-level in the spectra is $\sim$1.5~Jy. 

\subsection{Calibration}

The calibration of the intensities and the velocities is  one of the main concerns for variability studies.

The antenna
gain as a function of elevation is determined by daily observations of the
continuum source DR~21 (for which we assume a flux density of 16.4 Jy after
scaling the value of 17.04~Jy given by  Ott et al.~\cite{ott} for the ratio
of the source size to the 
Medicina beam\footnote{In previous works and up to
April 2003 the flux density of 18.8 Jy was adopted for DR~21, following Dent~(\cite{dent}). 
In this paper, {\bf all} data were recalibrated using the value given by
Ott et al.~(\cite{ott}).}) 
at a range of elevations.
Antenna temperatures were derived from total power measurements in position
switching mode. Integration time in each position is 10 sec with 400~MHz
bandwidth. The zenith system temperature is about 120 K in clear
weather conditions.

The daily gain curve was determined by fitting a polynomial curve to the data,
which was then used to convert antenna temperature to flux density for all
spectra taken that day. From the dispersion of the single measurements around
the curve we find the typical calibration uncertainty to be 19\%. On the few
days for which  no separate gain curve was measured, we applied that which was closest
in time, and estimated a corresponding calibration uncertainty of 7\%. The
average overall calibration uncertainty is therefore estimated to be $\sim$20\%.

To reduce the effects of atmospheric attenuation variations, observations
were obtained in good weather conditions and at elevations $\geq$ 30\degs, or  around
source meridian transit for low declination sources.  
For several sources we compared our intensities with those available in the
literature and close in time with our observations, 
virtually always finding very good agreement within the above-quoted
uncertainty (e.g., Valdettaro et al.~\cite{VPBCCFP2002}).
A good agreement was also found comparing observations 
taken at Effelsberg in a program to study
\hdo\ masers associated with  late--type stars which was run jointly by the Medicina
and Effelsberg radiotelescopes. 

As for the calibration of the velocity scale and its stability
over time, checks were made
with quasi simultaneous Effelsberg observations showing a 
good agreement. However, it is difficult to quote a definite
uncertainty considering the variable nature of the sources and the not
simultaneous observations. To make sure that the velocity changes that will
be reported are real effects and not due to instrumental changes (e.g., local
oscillator drifts in our system) and to evaluate the uncertainty,
we performed internal consistency checks within the sources of our sample.
The source G32.74-0.08 is characterized by a single, narrow and intense velocity
component. The observed velocity of the peak of the emission displays a maximum deviation from
a linear fit over the entire period of $<$ 0.1 \kms, a value that will be used
as maximum possible uncertainty on the velocity throughout our patrol. 

\section{The database}

For the purposes of the present study, the standard reduction procedures
were applied to the observed  spectra before including them  in the database:
\begin{itemize}
\item [(a)]
the edges of the available band were deleted to limit the spectra to 
the  flatter part of the band. This means that the velocity
coverage shown may be smaller (generally by 20-30\%) than the nominal bandwidth of 
the autocorrelator;
\item [(b)]
the spectra taken within 10 days were averaged, unless a significant
difference was found;
\item [(c)]
a polynomial  fit to the channels free of line emission was removed from the spectra;
\item [(d)] 
rebinning to 0.3 \kms\ was applied.
\end{itemize}

The whole set of Medicina 32--m observations
is archived in CLASS\footnote{CLASS is part of the GILDAS software developed at IRAM 
and Observatoire de Grenoble (http://www.iram.fr/IRAMFR/GILDAS)}  format. 
The raw data can be accessed through contact with
the authors.

The amount of information present in the database is so large that  
a concise way of presentation is essential for a rapid visual appreciation 
of the properties of maser variability. Below we describe our selection. 
All plots and figures presented can be found in electronic form
in the online appendix.

\subsection{The spectra}

The most obvious presentation is a  time sequence of
spectra. These are arranged in a compact form as page-plots in which each page 
contains 18 spectra, with time running from top to bottom and from left 
to right. In each spectrum, the date of observation is reported in the
top left corner, and the number of days elapsed since the first observation
of the source is shown in the top right corner.
The velocity scale is the same for all spectra of
each  source and is bracketed  by the minimum and maximum velocities at which
emission has been detected (above the 5 $\sigma$ level) throughout the 
monitoring period.
Given the large range of intensities present during our patrol,
we present the spectra using an autoscaled flux density scale.
These plots are convenient to view the full details of each 
spectrum except for those with bursts, although they have the disadvantage that the
flux density scale
changes from spectrum to spectrum and the visual impression of the variability is lost.

\begin{figure*}[tp]
\resizebox{14cm}{!}{
        \rotatebox{0}{
                \includegraphics{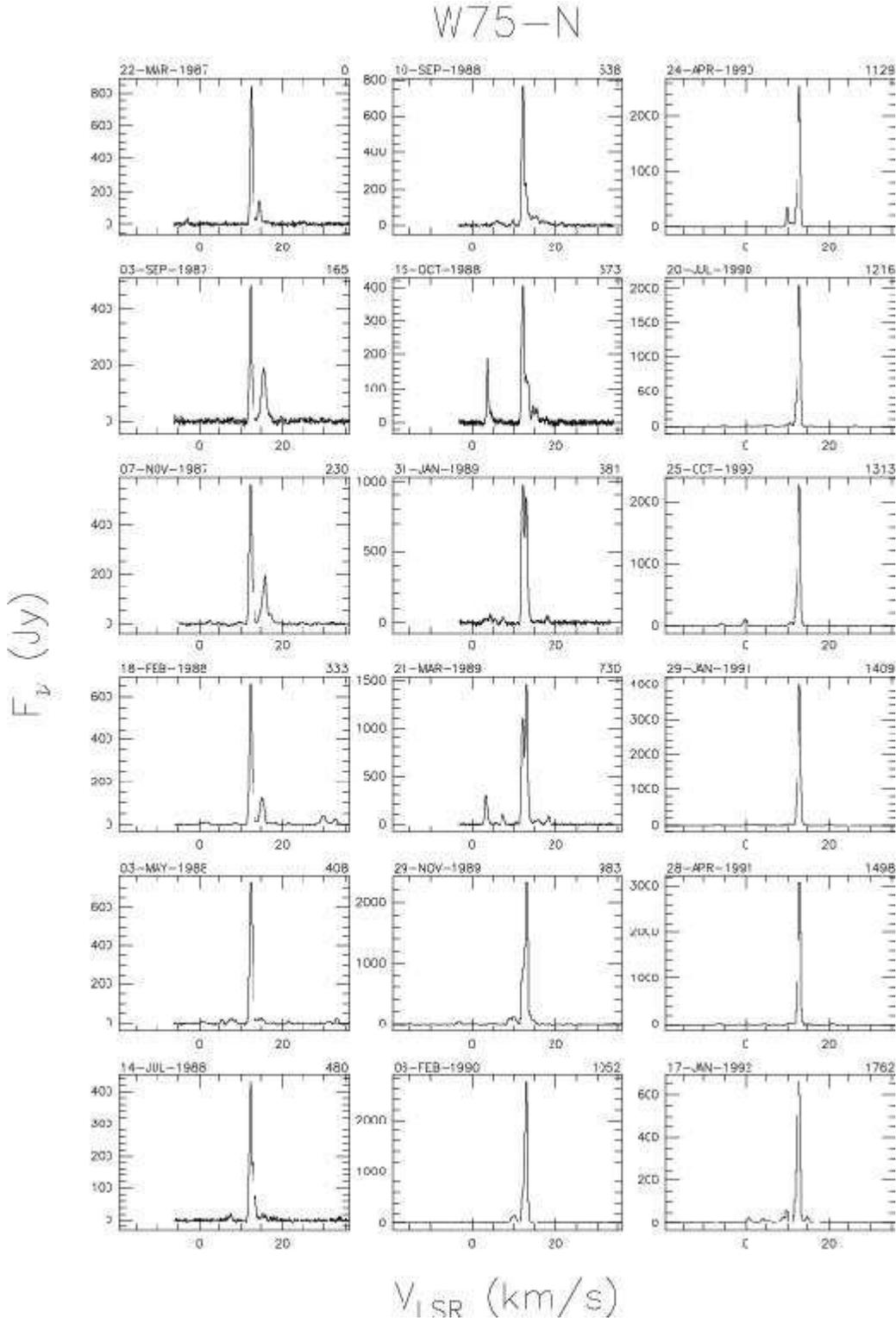}
        } }
\caption[]{
The first 18 spectra of the source W75--N with an 
autoscaled flux density scale.
Time runs from top to bottom and from left
to right. In each spectrum the actual date of observation is given above the
top left corner, and the number of days elapsed since the first observation
of the source above the top right corner. The velocity scale is the same for 
all spectra.
}
\label{w75n_tot_plots_01.ps}
\end{figure*}

As an example we show in Fig.~\ref{w75n_tot_plots_01.ps} the first page of
the spectra for the source W75--N.
 
The following two additional presentations of the spectra can be 
downloaded from our WEB pages\footnote{ 
http://www.arcetri.astro.it/$\sim$starform/water\_maser\_v2.html,
http://www.ira.inaf.it/papers/masers/water\_maser\_v2.html}.
\begin{itemize}
\item [(1)] Fixed--linear flux density scale, where the maximum
flux density is determined by the maximum value observed
during our monitoring. These plots are convenient for a quick
estimate of the variability, but compress the spectra too much
when the source presents bursts with intensities orders of
magnitude larger than the ``quiescent'' value.
\item [(2)] Same as in (1) but with logarithmic flux density scale.
The minimum
flux density is equal to the 5 $\sigma$ level in each spectrum.
\end{itemize}

\subsection{The velocity--time--flux density plots}

A more concise visualization of the time variation of
the maser emission can be obtained by showing the flux density versus
velocity and time in the same diagram. 

Since the observed dates are not distributed evenly in time, some kind of
interpolation between two adjacent observations must be used.  As no
assumption can be made on the evolution of the \hdo\ maser components, we
chose  a linear interpolation. This procedure produces an apparent  
increase in the lifetime of a feature when long time-intervals 
between two consecutive observations occur.  

Two versions of these plots are presented:
\begin{itemize}
\item [(1)] {\it full} plots, with velocity coverage extending to the maximum
velocity covered during the observations;
\item [(2)] {\it zoomed}  plots, with velocity coverage limited to the 
part of the spectrum where  emission has been detected (above the 5 $\sigma$ level)
at least once during the patrol.
\end{itemize}

In these plots, only emission above the 5 $\sigma$ level is shown.
In both plots the vertical solid line indicates the velocity of the
thermal molecular gas (either CO, CS or NH$_3$). In the {\it zoomed}
plots the additional dashed line represents the mean velocity derived from the 
histogram of the rate--of--occurrence (see Sect. 4.5). The time-scale on the left is in days
starting from the first observation. To convert this scale to 
real dates one should use the date given in the first spectrum 
of the figure with the time sequence of spectra, e.g.,
Fig.~\ref{w75n_tot_plots_01.ps} for W75--N or the date listed in
Table~1.
These plots are
particularly useful to find  possible velocity drifts and
to separate steady components from bursts of short duration.

\begin{figure*}[tp]
\resizebox{\hsize}{!}{
        \rotatebox{270}{
                \includegraphics{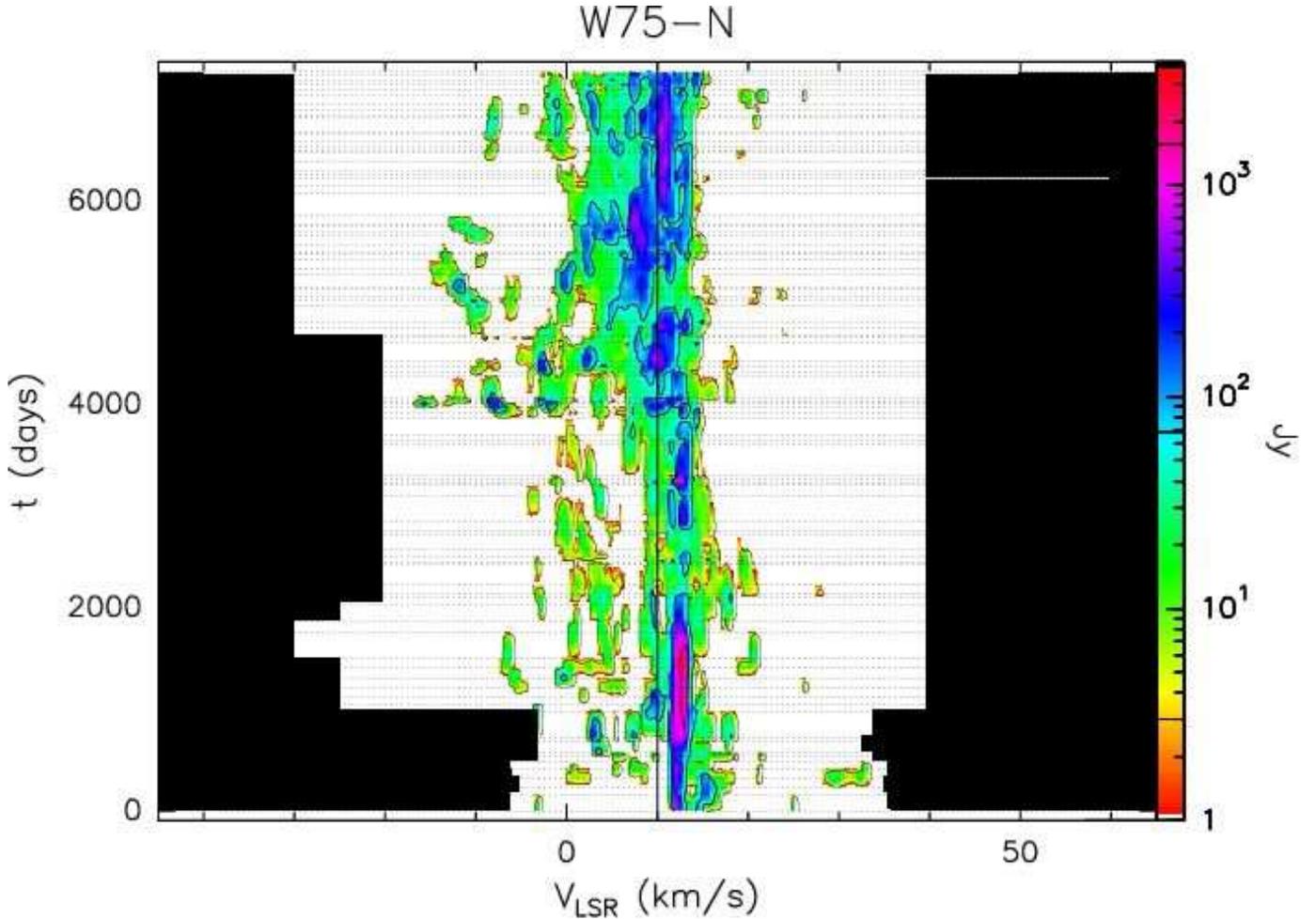}
        } }
\caption[]{The velocity--time--flux density {\it full} plot of the
source W75--N. The vertical solid line indicates the velocity
of the associated thermal molecular gas. The flux density scale
is shown by the bar on the right. In this bar the three lines
give the flux density of the drawn contours. The area in black 
indicates that the corresponding velocity range was not observed.
(This figure is in colour both in the electronic
 edition and in the online appendix.)
}
\label{w75n_grey_full.ps}
\end{figure*}

\begin{figure*}[tp]
\resizebox{\hsize}{!}{
        \rotatebox{270}{
                \includegraphics{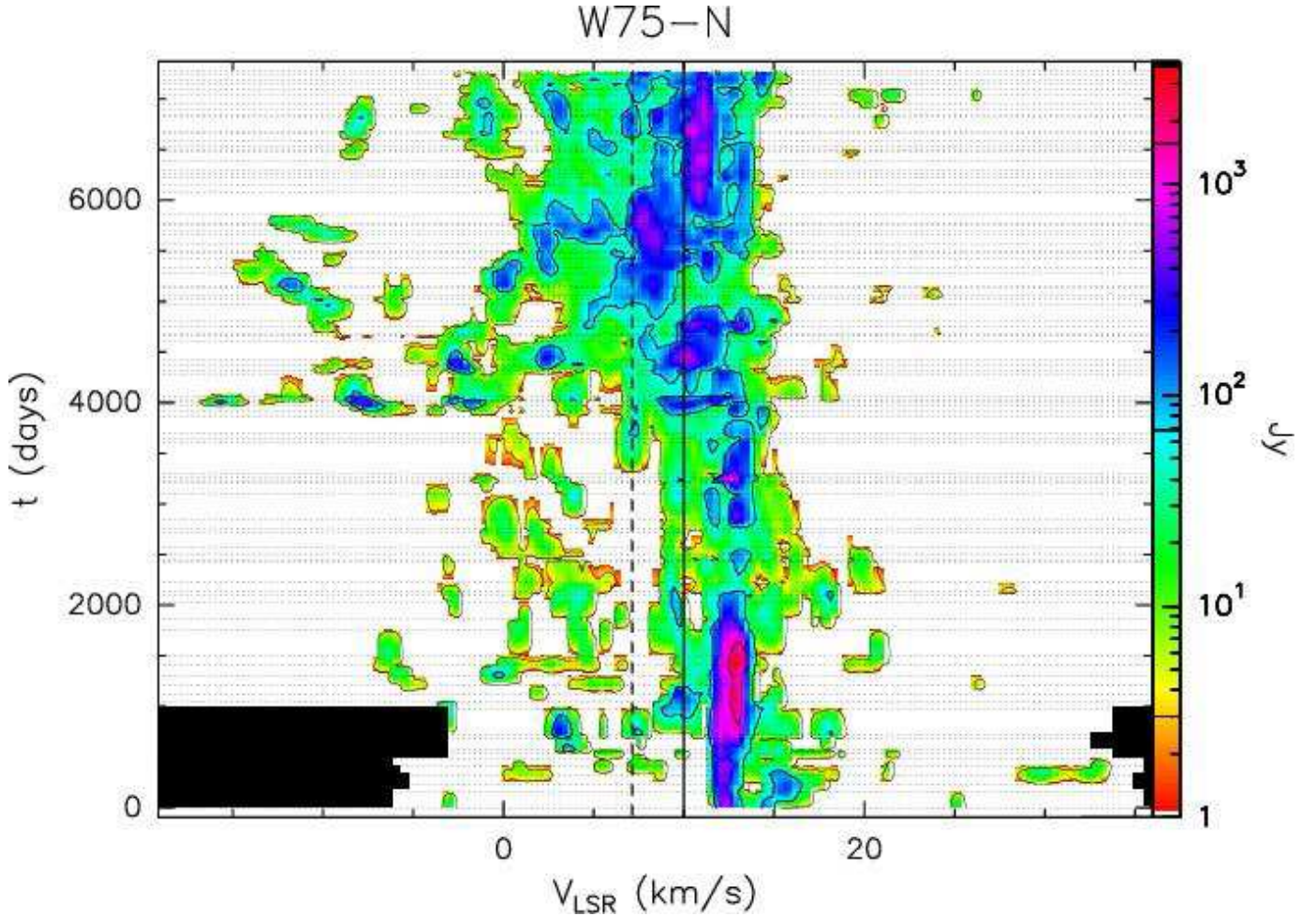}
        } }
\caption[]{The velocity--time--flux density {\it zoomed} plot of the
source W75--N. The vertical solid line indicates the velocity
of the associated thermal molecular gas. The  
vertical dashed line marks the mean velocity derived from the
histogram of the rate--of--occurrence. The flux density  scale
is shown by the bar on the right. In this bar the three lines
give the flux density of the drawn contours.
%
%
The area in black
indicates that the corresponding velocity range was not observed.
%
%
(This figure is in colour both in the electronic
 edition and in the online appendix.)
}
\label{w75n_grey_zoom.ps}
\end{figure*}

As an example of the velocity--time--flux density {\it full} and {\it zoomed} plots, 
in Fig.~\ref{w75n_grey_full.ps}--\ref{w75n_grey_zoom.ps}
we show those of the source W75--N.

\subsection{The light curve of the maser emission}

A plot of the peak flux density as a function of time 
is often used to study  
the temporal behaviour of the maser emission. However, the peak flux density is meaningful  
only if one narrow component is present throughout the entire observing period.  It
has less physical meaning for complex spectra with multiple components of comparable
flux densities varying in an uncorrelated fashion.  
In this case, the integrated flux density can only be used 
with the understanding that this represents the contribution of {\it all}
components and may be representative of the energy input from the YSOs.
However, since each velocity component may vary independently, 
the relative contribution of each to the light curve is lost in this plot.

\begin{figure}[tp]
\resizebox{\hsize}{!}{
        \rotatebox{270}{
                \includegraphics{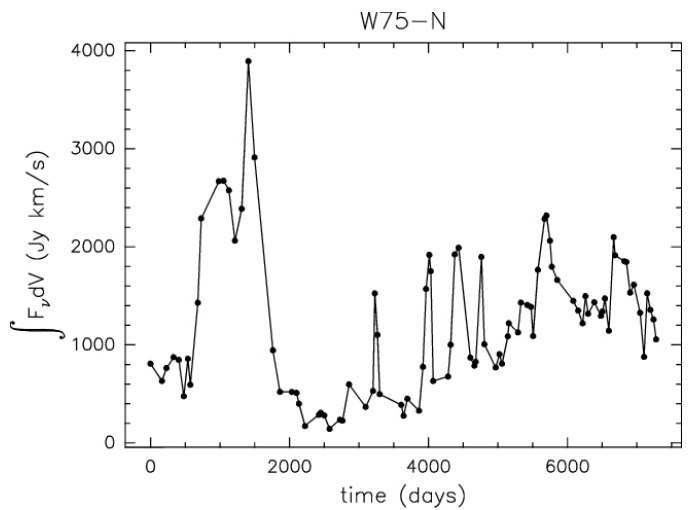}
        } }
\caption[]{The integral of the  flux density over the observed velocity range
as a function of time for  the
source W75--N.
}
\label{w75n_flux.ps}
\end{figure}

An example of the light curve of the integrated flux density for  the 
source W75--N is shown in Fig.~\ref{w75n_flux.ps}.

\subsection{The upper and lower envelopes and the mean spectrum}

Another meaningful description of  the degree of maser variability
is given by the comparison between the upper and lower envelopes 
and the mean spectra of
the sources over the whole period of observation.  

The upper envelope corresponds to the maximum
flux density ever reached at each velocity and is obtained by finding
the maximum flux density in each velocity bin. 
Its integral and the derived
luminosity, $L_{\rm H_2O}$(up), represent the maximum emission that could be
produced by the source  if {\it all} the velocity components were to emit at their
{\it maximum} level and at the {\it same} time. 

The mean spectrum represents the mean flux density at each velocity, 
and is obtained by computing the arithmetic mean of the flux
densities in each velocity bin, assigning the same weight to the individual spectra. 
In this average the
flux density is arbitrarily set to zero if it is below the
5 $\sigma$ noise level in the spectrum in question.  Clearly, the steadier components 
dominate this spectrum.  

Finally, the lower envelope
identifies the components that are always active and is obtained by finding
the minimum flux density in each velocity bin, again setting it to zero if it is below the
5 $\sigma$ noise level in each  spectrum. When the lower envelope is
zero over the whole velocity interval, it indicates that no velocity
component is constantly present above a 5 $\sigma$ level during 
the monitoring  period. This does not
necessarily imply that the maser is quiescent at all velocities at some epoch, 
since the emission may occur at different velocities  at different times.

\begin{figure}[tp]
\resizebox{\hsize}{!}{
        \rotatebox{270}{
                \includegraphics{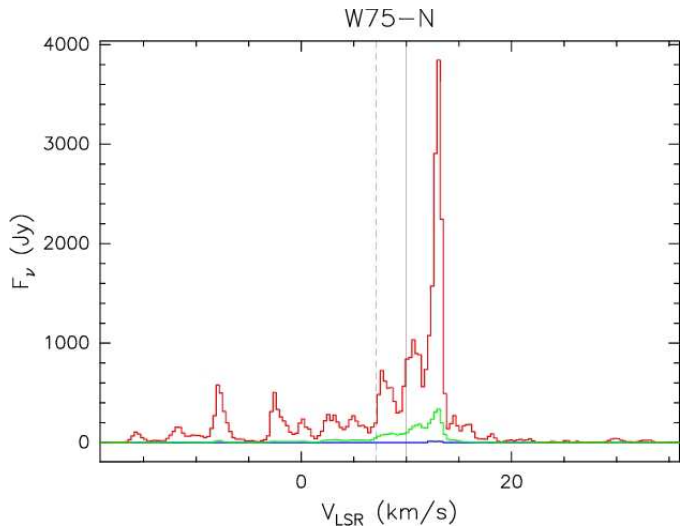}
        } }
\caption[]{The upper (red) and  lower (blue)  envelopes and the mean spectrum (green)
of the source W75--N measured during our monitoring. The vertical solid line marks 
the velocity of the associated thermal molecular gas. The vertical dashed  line
marks the mean velocity derived from the
histogram of the rate--of--occurrence.
(This figure is in colour both in the electronic
 edition and in the online appendix.)
}
\label{w75n_uplo.ps}
\end{figure}

An example of the upper and lower envelopes, and mean spectrum of the
source W75--N is shown in Fig.~\ref{w75n_uplo.ps}.

\subsection{The rate--of--occurrence plots}
 
To identify the velocity components which are more frequently
present in the spectra (independently of their flux density), we plot the
rate--of--occurrence of maser emission above the 5 $\sigma$ noise level as a
function of  velocity. To produce this plot we have 
set a counter for each channel that increases by one unit every time the flux
density in the channel is greater than the 5$\sigma$ noise level of the spectrum.

\begin{figure}[tp]
\resizebox{\hsize}{!}{
        \rotatebox{270}{
                \includegraphics{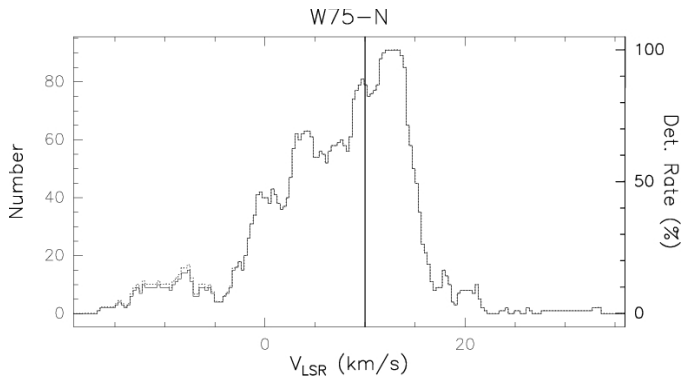}
        } }
\caption[]{The rate--of--occurrence plot for the
source W75--N. The scale to the right refers to
the dotted histogram, the scale to the left to the solid line  
histogram.  The vertical solid line marks 
the velocity of the associated thermal molecular gas.
}
\label{w75n_histo.ps}
\end{figure}

An example of the rate--of--occurrence  histogram
of the source W75--N is shown in Fig.~\ref{w75n_histo.ps}.

Two histograms can  be produced:  

\begin{itemize}
\item [(1)] the percentage of detections with respect to the total number of observations 
(dotted line histogram, scale to the right); 
\item [(2)] the number of detections (solid line histogram, scale to the left).
\end{itemize}

The two histograms may differ slightly in velocity ranges where the time coverage is
not equal for all channels, produced by using different bandwidths during
our monitoring. 
From this histogram we can derive the mean velocity which is shown in Figs.~\ref{w75n_grey_zoom.ps} 
and~\ref{w75n_uplo.ps}.

\section{Conclusions}
We present the observational results of a systematic study 
extending for almost 20 years of
the \hdo\ maser variability in 43 SFRs with luminosities
of the associated IR sources between 20 and 1.5$\times 10^6$ L$_\odot$.
This database provides the backbone for the discussion of the main
long-term properties of maser emission that will be presented 
in a forthcoming paper. We have identified several
ways to describe graphically the main aspects of the \hdo\ maser 
emission. These include:

\begin{itemize}
\item [(a)] the spectra in a compressed form, with 
an autoscaled  flux density scales 
(the same sets of plots,
but with fixed linear and logarithmic
flux density scales, are also available from our WEB pages);
\item [(b)] the velocity--time--flux density  plots, which  conveniently describe
the morphology of the variability of the maser emission. These diagrams are  
particularly useful for recognizing the presence of possible velocity drifts and
separating steady components from bursts of short duration; 
\item [(c)] the velocity--integrated flux density as a function of time. 
This light curve   describes the overall
emission  of the maser components associated with a SFR;  
\item [(d)] the mean spectrum; 
\item [(e)] the upper and lower  envelopes
of the maser emission. The upper envelope  
represents the maximum emission that the SFR could
produce if {\it all} the velocity components were present {\it simultaneously}
and emitting at their {\it maximum} rate. Similarly, the lower envelope pinpoints
the steady components and their lowest level  (but $> 5 \sigma$) of emission;
\item [(f)] the number of times that  the maser emission  is well above the noise
($> 5 \sigma$) as a function of 
velocity is  directly gauged by the histogram of the rate--of--occurrence.
\end{itemize}

\begin{acknowledgements}
This long term project could not have been carried out without the help
and dedication of the technical staff of the Radio Group of the Arcetri 
Observatory and the constant assistance of the technical staff of the Medicina 32--m 
Radiotelescope. This research has made use of the SIMBAD database, operated at CDS,
Strasbourg, France.
\end{acknowledgements}


\begin{thebibliography}{}
\bibitem[1982]{antw} 
  Anthony--Twarog, B.\ J.\ 1982, \aj, 87, 1213
\bibitem[1981]{arma} 
  Armandroff, T.\ E., \& Herbst, W.\ 1981, \aj, 86, 1923
\bibitem[1959]{bla59} 
  Blaauw, A., Hiltner, W.\ A., \& Johnson, H.\ L.\ 1959, \apj, 130, 69
\bibitem[1991]{bos}
  Bos, A.\ 1991, IEEE Trans. on Instrumentation and Measurements, 40, 591
\bibitem[1993]{bb93}
  Brand, J., \& Blitz, L.\ 1993, \aap, 275, 67
\bibitem[1994]{brand}
  Brand, J., Cesaroni, R., Caselli, P., et al.\ 1994, \aaps, 103, 541
\bibitem[1998]{bw98} 
  Brand, J., \& Wouterloot, J.\ G.\ A.\ 1998, \aap, 337, 539
\bibitem[2003]{BCCFPPV2003}
  Brand, J., Cesaroni, R., Comoretto, G., et al.\ 2003, \aap, 407, 573
\bibitem[2007]{brand2007}
  Brand, J., Felli, M., Cesaroni, R., et al.\ 2007, in Astrophysical Masers and Their Environments,
  Proceedings of IAU Symposium No 242, ed.\ J.\ Chapman, \& W.\ A.\ Baan, in Press
\bibitem[1988]{cesa88}
  Cesaroni, R., Palagi, F., Felli, M.,  et al.\ 1988, \aaps, 76, 445
\bibitem[1999]{CFW98}
  Cesaroni, R., Felli, M., \& Walmsley, C.\ M.\  1999, \aaps, 136, 333 
\bibitem[1990]{cwc} 
  Churchwell, E., Walmsley, C.\ M., \& Cesaroni, R.\ 1990, \aaps, 83, 119
\bibitem[2006]{clarke}
  Clarke, A.\ J., Lumsden, S.\ L., Oudmaijer, R.\ D., et al.\ 2006, \aap, 457, 183
\bibitem[1996]{cla96}
  Claussen, M.\ J., Wilking, B.\ A., Benson, P.\ J., et al.\ 1996, \apjs, 106, 111
\bibitem[1994]{CFNPP94}
  Codella, C., Felli, M., Natale, V., Palagi, F., \& Palla, F.\  1994, \aap, 291, 261
\bibitem[1995]{CF95}
  Codella, C., \& Felli, M.\ 1995, \aap, 302, 521
\bibitem[1995]{CP95}
  Codella, C., \& Palla, F.\ 1995, \aap, 302, 528 
\bibitem[1996]{CFN96}
  Codella, C., Felli, M., \& Natale, V.\ 1996,  \aap, 311, 971
\bibitem[1990]{com90}
  Comoretto, G., Palagi, F., Cesaroni, R., et al.\ 1990, \aaps, 84, 179
\bibitem[1974]{cramfis} 
  Crampton, D., \& Fisher, W.\ A.\ 1974, Pub. Dom.  Astrophys. Obs., 14, 283
\bibitem[1972]{dent}
  Dent, W.\ A.\ 1972, \apj, 177, 93
\bibitem[1999]{deze} 
  de Zeeuw, P.\ T., Hoogerwerf, R., de Bruijne, J.\ H.\ J., Brown, A.\ G.\ A.,
  \& Blaauw, A.\ 1999, \aj, 117, 354
\bibitem[1981]{nayo} 
  Evans, N.\ J., \& Blair, G.\ N.\ 1981, \apj, 246, 394
\bibitem[1992]{FPT92}
  Felli, M., Palagi, F., \& Tofani, G.\ 1992, \aap, 255, 293
\bibitem[2006]{FMRC2006}
  Felli, M., Massi, F., Robberto, M., \& Cesaroni, R.\ 2006, \aap, 453, 911
\bibitem[1989]{for89} 
  Forster, J.\ R., \& Caswell, J.\ L.\ 1989, \aap, 213, 339
\bibitem[1999]{for99} 
  Forster, J.\ R., \& Caswell, J.\ L. 1999, \aaps, 137, 43
\bibitem[1986]{fukui} 
  Fukui, Y., Sugitani, K., Takaba, H., et al.\ 1986, \apj, 311, L85
%
%
\bibitem[2003]{furuya} 
  Furuya, R.\ S., Kitamura, Y., Wootten, A., Claussen, M.\ J., \&
  Kawabe, R.\ 2003, \apjs, 144, 71 (Erratum: 2007, ApJS, 171, 349)
%
%
\bibitem[1975]{geo75} 
  Georgelin, Y.\ 1975, Th\`{e}se de Doctorat, Universit\'{e} de Provence
\bibitem[2006]{hachi06} 
  Hachisuka, K., Brunthaler, A., Menten, K.\ M., et al.\ 2006, \apj, 645, 337
\bibitem[1983]{hj83}
  Herbig, G.\ H., \& Jones, B.\ F.\ 1983, \aj, 88, 1040
\bibitem[2005]{hon05} 
  Honma, M., Bushimata, T., Choi, Y. K., et al.\ 2005, PASJ, 57, 595
\bibitem[1990]{massey} 
  Hunter, D.\ A., \& Massey, P.\ 1990, \aj, 99, 846
\bibitem[1994]{hun94}
  Hunter, T.\ R., Taylor, G.\ B., Felli, M., \& Tofani, G.\  1994, \aap, 284, 215
\bibitem[2007]{jeff} 
  Jeffries, R.\ D.\ 2007, \mnras, 376, 1109
\bibitem[1979]{lawo} 
  Lada, C.\ J., \& Wooden, D.\ 1979, \apj, 232, 158
\bibitem[1989]{lil89}
  Liljestr\"om, T., Mattila, K., Toriseva, M., \& Anttila, R.\ 1989, \aaps, 79, 19
\bibitem[1977]{lit77}
  Little L.\ T., White, G.\ J., \&  Riley, P.\ W.\ 1977, \mnras, 180, 639 
\bibitem[1998]{meeha}
  Meehan, L.\ S.\ G., Wilking, B.\ A., Claussen, M.\ J., Mundy, L.\ G., \& 
  Wootten, A.\ 1998, \aj, 115, 1599
\bibitem[1996]{mol96} 
  Molinari, S., Brand, J., Cesaroni, R., \& Palla, F.\ 1996, \aap, 308, 573
\bibitem[2002]{mol02} 
  Molinari, S., Testi, L., Rodr\'{i}guez, L.\ F., \& Zhang, Q.\ 2002, \apj, 570, 758
\bibitem[1994]{ott}
  Ott, M., Witzel, A., Quirrenbach, A., et al.\ 1994, \aap, 284, 331
\bibitem[1993]{PCC93}
  Palagi, F., Cesaroni, R., Comoretto, G., Felli, M., \& Natale, V.\ 1993, \aaps, 101, 153
\bibitem[1991]{pal91}
  Palla, F., Brand, J., Comoretto, G., Felli, M., \& Cesaroni, R.\ 1991, \aap, 246,  249
\bibitem[1993]{PCBCCF93}
  Palla, F., Cesaroni, R., Brand, J., et al.\ 1993, \aap, 280, 599
\bibitem[1994]{per94}
  Persi, P., Palagi, F., \& Felli, M.\ 1994, \aap, 291, 577
\bibitem[1968]{raci}
  Racine, R.\ 1968, \aj, 73, 233
\bibitem[1970]{ravdb} 
  Racine, R., \& van den Bergh, S.\ 1970, Reflection Nebulae and Spiral Structure, 
  in The Spiral Structure of our Galaxy, ed.\  W.\ Becker, \& G.\ I.\ Kontopoulos 
  (Dordrecht: Reidel), IAU Symp.\ 38, 219
\bibitem[2007]{russi}
  Rudnitskij, G.\ M., Paschenko, M.\ I., Lekht, E.\ E., et al.\ 2007, in 
  Astrophysical Masers and Their Environments, Proceedings of IAU Symposium No 242, 
  ed.\ J.\ Chapman, \& W.\ A.\ Baan, in Press
\bibitem[1996]{devi} 
  Shepherd, D.\ S., \& Churchwell, E.\ 1996, \apj, 472, 225
\bibitem[1988]{sn88} 
  Snell, R.\ L., Huang, Y.-L., Dickman, R.\ L., \& Claussen, M.\ J.\ 1988, \apj, 325, 853
\bibitem[1995]{tof95}
  Tofani, G., Felli, M., Taylor, G.\ B., \& Hunter, T.\ R.\ 1995, \aaps, 112, 299
\bibitem[2001]{val01}
  Valdettaro, R., Palla, F., Brand, J., et al. 2001, A\&A, 368, 845
\bibitem[2002]{VPBCCFP2002}
  Valdettaro, R., Palla, F., Brand, J., et al. 2002, A\&A, 383, 244
\bibitem[1999]{vdt} 
  van der Tak, F.\ F.\ S., van Dishoeck, E.\ F., Evans, N.\ J., Bakker, E.\ J.,
  \& Blake, G.\ A.\ 1999, \apj, 522, 991
\bibitem[1956]{walker} 
  Walker, M.\ F.\ 1956, \apjs, 2, 365
\bibitem[1989]{wilky} 
  Wilking, B.\ A., Blackwell, J.\ H., Mundy, L.\ G., \& Howe, J.\ E.\ 1989, \apj, 345, 257
\bibitem[1989]{WC89}
  Wood, D.\ O.\ S., \&  Churchwell, E.\ 1989, \apj, 340, 265
\bibitem[1986]{wouw} 
  Wouterloot, J.\ G.\ A., \& Walmsley, C.\ M.\ 1986, \aap, 168, 237
\bibitem[1995]{wou95}
  Wouterloot, J.\ G.\ A., Fiegle, K., Brand, J., \& Winewisser, G.\ 1995, 
  \aap, 301, 236 (Erratum: 1997, \aap, 319, 360)
\end{thebibliography}
\end{document}